\documentclass[12pt]{article}
\usepackage{graphicx}

\def\lsim{\;
\raise0.3ex\hbox{$<$\kern-0.75em\raise-1.1ex\hbox{$\sim$}}\;
}

\def\gsim{\;
\raise0.3ex\hbox{$>$\kern-0.75em\raise-1.1ex\hbox{$\sim$}}\;
}

\begin{document}

\title{{\bf \large QCD SPECTRAL SUM RULES AND
SPONTANEOUSLY BROKEN CHIRAL SYMMETRY} \thanks{Work supported in part
by BMBF, GSI and Conselleria de Cultura, Educaci\'o i Ci\`encia de la
Generalitat Valenciana.}
}

\author{E. Marco and W. Weise\\
\\
Physik-Department,\\
Technische Universit\"at M\"unchen,\\
D-85747 Garching, Germany}

\maketitle

\begin{abstract}
The gap $\Delta = 4 \pi f_{\pi} \simeq 1.2$ GeV of spontaneous chiral
symmetry breaking is introduced as a scale delineating resonance and
continuum regions in the QCD spectral sum rules for vector mesons. Basic
current algebra results are easily recovered, and accurate sum rules
for the lower moments of the spectral distributions are derived. The
in-medium scaling of vector meson masses finds a straightforward
interpretation, at least in the narrow width limit.
\end{abstract}


\newpage

Spectral sum rules have frequently been used to connect observable
information through dispersion relations with the operator product
expansion (OPE) in QCD, in the form of either SVZ sum rules
\cite{SVZ} or finite energy sum rules (FESR)
\cite{Chetyrkin, Fischer, Klingl}. Prototype
examples are the sum rules for the lightest ($\rho$ and $\omega$) vector
mesons. The starting point is the vacuum current-current
correlation function,

\begin{eqnarray}   \label{eq:corrfunc}
\Pi_{\mu \nu} (q) &=& i \int d^4x \, e^{i q \cdot x}
\langle 0 | {\cal T} j_{\mu} (x) j_{\nu} (0) | 0 \rangle\nonumber \\
&=&(g_{\mu \nu} - \frac{q_{\mu} q_{\nu}}{q^2})\Pi (q^2) \, ,
\end{eqnarray}
where ${\cal T}$ denotes the time-ordered product and
the currents are specified, for the case of interest here, as
$j_{\mu}^{(\rho)} = (\bar{u} \gamma_{\mu} u -
\bar{d} \gamma_{\mu} d)/2$ for the $\rho$ channel and
$j_{\mu}^{(\omega)} = (\bar{u} \gamma_{\mu} u +
\bar{d} \gamma_{\mu} d)/6$ for the $\omega$ channel. We work as usual with
the spectrum

\begin{equation}    \label{eq:espectral}
R(q^2) = - \frac{12\pi}{q^2} \mbox{Im} \Pi (q^2)
\end{equation}
normalized to the ratio $ \sigma(e^+ e^- \rightarrow \mbox{hadrons})/
\sigma(e^+ e^- \rightarrow \mu^+ \mu^-)$.

The nuclear physics interest in QCD sum rules is motivated by
applications of SVZ type Borel sum rules not only in vacuum, but
also in nuclear matter in order to extract in-medium properties
of vector mesons \cite{Hatsuda, KlinglKaiser}. The commonly adopted
procedure is to use a
schematic ``duality'' ansatz for $R$, with the vector meson resonance
represented as a $\delta$-function and a step function continuum starting
at a threshold $s_0$. The position of the resonance and the threshold
$s_0$ are then fitted by requiring consistency with the vacuum or
in-medium OPE side of the sum rule.

The continuum threshold $s_0$ is usually introduced as a free parameter.
On the other hand, spontaneous chiral symmetry breaking in QCD suggests
that the mass gap which separates the QCD ground state
from the high energy continuum should be expressed as some multiple
of the pion decay constant, $f_{\pi} = 92.4$ MeV, since this is the only
remaining scale in the limit of vanishing quark masses.

In the present paper we propose to identify this
continuum threshold with the chiral gap parameter,
$\Delta = 4 \pi f_{\pi} \simeq 1.2$ GeV, by setting

\begin{equation}       \label{eq:gap}
\sqrt{s_0} = \Delta = 4 \pi f_{\pi} \, ,
\end{equation}
and thereby unifying QCD sum rules with spontaneous chiral symmetry breaking.

Starting from this assumption we shall demonstrate that
QCD spectral sum rules combined with vector meson dominance
(VMD) immediately imply well known current algebra relations, a very
welcome feature. When realistic spectral distributions are used, the
transition between resonance region and continuum is no longer
sharp, but the chiral mass gap (\ref{eq:gap}) is shown still
to control the smooth turnover from the hadronic part to the asymptotic
QCD domain of the spectrum for both $\rho$ and $\omega$ channels. We also
point out briefly that our hypothesis (\ref{eq:gap}) permits one to
understand the in-medium results of ref.\ \cite{Hatsuda} in a simple and
straightforward way, using the leading density dependence of the pion
decay constant.

Our approach is based on rigorous sum rules
for the lowest moments of the spectral distribution (\ref{eq:espectral}).
A direct access to these moment sum rules is best given by the FESR
method \cite{Chetyrkin, Fischer, Gimenez}. Consider
the vacuum correlation function $\Pi (q^2 = s)$ of
Eq.\ (\ref{eq:corrfunc}) in the complex $s$-plane where it has a cut
along the positive real axis. Choose a closed loop $\gamma$ consisting of a
path which surrounds and excludes the cut along$\mbox{ Re}s > 0$, and joins
with a circle $C_{s_0}$ of fixed radius $s_0$. Cauchy's theorem implies
$\oint_{\gamma} ds \; s^{N-1} \Pi(s) = 0$ for integer $N \geq 0$.
Separating this integral and using Eq.\ (\ref{eq:espectral}) gives

\begin{equation}    \label{eq:loop}
\int_0^{s_0} ds \; s^N R(s) = - 6\pi i \oint_{C_{s_0}} ds \; s^{N-1}
\Pi(s) = 6 \pi s_0^N \int_0^{2 \pi} d\theta e^{i N \theta}
\Pi(s_0 e^{i \theta})\, .
\end{equation}
It remains to evaluate the r.h.s.\ integral along the circle of
radius $s_0$. For sufficiently large $s_0$ one can use perturbative
QCD and add non-perturbative corrections via the OPE:

\begin{equation}    \label{eq:Pi}
\Pi(s) = \Pi^{\mbox{\scriptsize pQCD}} (s) + \frac{d}{12 \pi^2}
\sum_{n\geq0} (-)^n \frac{c_{n+1}}{s^n} \, ,
\end{equation}
with $d= 3/2$ or $1/6$ for the $\rho$ or $\omega$ channels, respectively.
The parameters $c_n$ have dimension (mass)$^{2n}$, with 
$c_1 = - 3 (m_u^2 + m_d^2)$ and the dimension-4 condensates
$c_2 = (\pi^2/3) \langle (\alpha_s/\pi) G_{\mu \nu} G^{\mu \nu}\rangle
+4 \pi^2 \langle m_u \bar{u} u + m_d \bar{d} d \rangle$
which are reasonably well under control. The $c_3$ involves the
(much less certain) four-quark condensates. In practice $c_1$ is negligibly
small, and the quark condensate piece in $c_2$ can be dropped in
comparison with the gluon condensate term. As usual, we ignore logarithmic
corrections to the condensates.

The pQCD part of $\Pi(s)$ is calculated to third order on $\alpha_s$ using
the $\overline{MS}$ scheme \cite{Gorishny}. The result is

\begin{equation}     \label{eq:PipQCD}
\Pi^{\mbox{\scriptsize pQCD}} (s) = \frac{d}{12 \pi^2} \sum_{n = 0}^{3}
\left(\frac{\alpha_s(\mu^2)}{\pi}\right)^n \Pi^{(n)}(s;\mu^2)
\end{equation}
at a renormalization point $\mu^2$, with $\Pi^{(0)} = s[K_0
+\ln(-s/\mu^2)]$ and $\Pi^{(n)} = s[K_n
+\sum_{m=1}^n A_{m n} \ln^m(-s/\mu^2)]$ for $n = 1$, 2, 3.
The constants $K_n$ are irrelevant since they drop out in the loop
integral (\ref{eq:loop}). The relevant coefficients
(for $N_f = 3$ flavours) of the logarithmic terms are
$A_{1 1} = 1$, $A_{1 2} = 1.641$, $A_{2 2} = -1.125$, $A_{1 3} = -10.28$,
$A_{2 3} = -5.69$,
$A_{3 3} = 1.69$. The renormalization point can be chosen at $\mu^2 = s_0$.

Inserting $\Pi(s_0 e^{i \theta})$ from Eqs.\ (\ref{eq:Pi},\ref{eq:PipQCD})
and using $\ln (-e^{i \theta}) = i (\theta - \pi)$, the r.h.s.\ integral of
Eq.\ (\ref{eq:loop}) is easily worked out and one arrives at the
following set of sum rules for the lowest spectral moments with
$N = 0$, 1, 2:

\begin{equation}     \label{eq:moments}
\int_0^{s_0} ds \; s^N R(s) = d\left[\frac{s_0^{N+1}}{N+1} (1 + \delta_N)
+ (-)^N c_{N+1}\right] \, .
\end{equation}
The perturbative QCD corrections up to $O(\alpha_s^3)$ are summarized as

\begin{eqnarray}     \label{eq:deltan}
\delta_N &=&\frac{\alpha_s}{\pi} + \left(\frac{\alpha_s}{\pi}\right)^2
\left[A_{1 2} - \frac{2}{N+1} A_{2 2}\right] \nonumber \\
 &&+ \left(\frac{\alpha_s}{\pi}\right)^3 \left[A_{13} - \frac{2}{N+1} A_{23}
+ \left(\frac{6}{(N+1)^2} - \pi^2\right) A_{33}\right] \, ,
\end{eqnarray}
where $\alpha_s$ is taken at $\mu^2 = s_0$. Note that different
$\delta_N$ apply for the various moments of $R(s)$, and that condensates
of different mass dimension appear well separated in the different moments.
For example, uncertain four-quark condensates enter only at $N=2$, whereas
the moments with $N=0,1$ are free of such uncertainties. It can be
readily demonstrated that the results, Eqs.\
(\ref{eq:moments},\ref{eq:deltan}), are rigorously consistent with
those obtained using the Borel sum rule method. It is also interesting
to note that our deduction of the sum rules (\ref{eq:moments}) is
analogous to the procedure used to extract $\alpha_s$ from $\tau$
decays \cite{Pich}.

Before applying these sum rules to realistic spectral distributions, we
turn first to a schematic model for $R(s)$ which combines vector
meson dominance (VMD) and the QCD continuum:

\begin{equation}     \label{eq:Rv}
R_V(s)= 12 \pi^2 \frac{m_V^2}{g_V^2} \delta(s - m_V^2) + \mbox{continuum}
\quad (V=\rho, \omega)\, ,
\end{equation}
with $g_{\rho} = g_{\omega}/3 = g$, the universal vector coupling
constant. For convenience we discuss the $\rho$ meson sum rules first.
We set $s_0 = 16 \pi^2 f_{\pi}^2$ according to our conjecture (\ref{eq:gap})
and find the sum rules for the first two moments:

\begin{equation}     \label{eq:VMD0th}
\int_0^{s_0} ds R_{\rho} (s) = 12 \pi^2 \frac{m_{\rho}^2}{g^2} = \frac{3}{2}
(4 \pi f_{\pi})^2 (1+\delta_0) + \frac{3}{2} c_1 \, ,
\end{equation}

\begin{equation}     \label{eq:VMD1st}
\int_0^{s_0} ds \; s R_{\rho} (s) = 12 \pi^2 \frac{m_{\rho}^4}{g^2} 
= \frac{3}{4}
(4 \pi f_{\pi})^4 (1+\delta_1) - \frac{3}{2} c_2 \, .
\end{equation}
Once the hypothesis (\ref{eq:gap}) is launched, there are no
free parameters in these sum rules. Dropping the QCD corrections for
the moment, the sum
rule (\ref{eq:VMD0th}) immediately gives

\begin{equation}     \label{eq:KSFR}
  m_{\rho}^2 = 2 g^2 f_{\pi}^2 \, ,
\end{equation}
the well-known KSFR relation \cite{KSFR}, while the sum rule (\ref{eq:VMD1st})
for the first moment further specifies

\begin{equation}     \label{eq:g2pi}
g = 2 \pi \, .
\end{equation}

The results (\ref{eq:KSFR},\ref{eq:g2pi}) are quite remarkable:
identifying the onset of the continuum spectrum with the gap
$\Delta = 4 \pi f_{\pi}$, the scale for spontaneous chiral
symmetry breaking, a unification of QCD spectral sum rules
with current algebra emerges, yielding $m_V = \sqrt{8}
\pi f_{\pi} = \Delta/\sqrt{2}$ in leading order (identical relations
hold for both $\rho$ and $\omega$ meson). The condition $g = 2 \pi$
is actually consistent with the effective action of the $SU(3)\times
SU(3)$ non-linear sigma model and the Wess-Zumino term. Application of the QCD
corrections moves $g$ to within less than 10\% of the empirical
$g_{\rho} \simeq 5.04$ deduced from the $\rho \rightarrow e^+ e^-$ decay width.

Let us now turn from schematic to realistic spectral distributions.
The vector mesons have energy dependent widths from their leading
decay channels $\rho \rightarrow \pi \pi$ and $\omega \rightarrow 3 \pi$,
etc. Multipion ($n\pi$) channels (with $n$ odd/even for $I=0,1$)
open up and continue toward the asymptotic pQCD spectrum. It would
then seem difficult, at first sight, to locate the gap $\Delta$ which
delineates the resonance from the QCD continuum. Remarkably, though, the
turnover from resonance to continuum is still governed by the scale
set by the chiral gap $\Delta = 4 \pi f_{\pi}$. One can simply
replace the sharp edge at $s_0 = \Delta^2$ by a smooth interpolation
in the interval $\Delta^2 - 0.6 \mbox{GeV}^2 \lsim s \lsim
\Delta^2 + 0.6 \mbox{GeV}^2$. In essence, the spreadings of the resonance
and the gap edge amount to incorporating $1/N_c$ corrections to the zero
width spectrum (\ref{eq:Rv}).

\begin{figure}[t]
\centerline{
\includegraphics[width=0.6\textwidth,angle=-90]{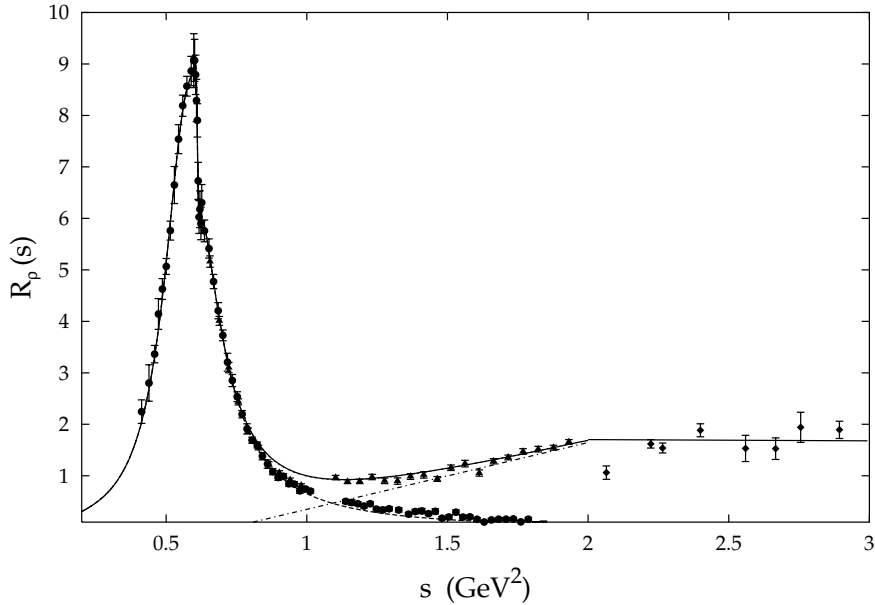}
}
\caption{Spectral distribution $R_{\rho}(s)$ in the isovector
($\rho$ meson) channel.
Dashed line: resonant part; dot-dashed line: linear interpolation centred
at $\Delta^2 = (4 \pi f_{\pi})^2$; solid line: sum of resonance
and continuum contributions. Experimental data: full dots
from $e^+ e^- \rightarrow \pi^+ \pi^-$ \cite{Barkov};
triangles \cite{Dolinsky} and diamonds \cite{Bacci} from total
$e^+ e^- \rightarrow n\pi$ with $n$ even.}
\label{fig1}
\end{figure}

In practice the calculation proceeds as follows. Consider the $\rho$ meson
first. The resonant part of the spectrum is well described using effective
field theory as shown in ref.\ \cite{KlinglKaiser}, including
$\rho \omega$ mixing. Guided by this approach we use a parametrized form,
given in Ref.\ \cite{Benayoun},
which reproduces the $e^+ e^- \rightarrow \pi^+ \pi^-$ data (see 
dashed curve in Fig.\ 1). We let the QCD
continuum start at $s_c = 2$ GeV$^2$ and use a linear
interpolation across the interval $0.8 \mbox{GeV}^2 \lsim s \lsim s_c$
centered at $\Delta^2$ (dashed-dotted curve in Fig.\ 1). When added to
the tail of the $\rho$ resonance this interpolation
obviously works well in reproducing
the total $e^+ e^- \rightarrow 2\pi, 4\pi,\ldots$ data in the $I=1$ channel.
We now employ the sum rules (\ref{eq:moments}) with $s_0=s_c$ and check
overall consistency, using $\alpha_s(s_0) = 0.39$.
For the lowest moment with $N=0$ the left-hand side gives
$\int_0^{s_c} ds \, R_{\rho}(s) = 3.527$ GeV$^2$, while the right-hand side
gives 3.521 GeV$^2$. For $N=1$ the l.h.s.\ integral gives 
$\int_0^{s_c} ds \, s R_{\rho}(s) = 3.27$ GeV$^4$, while the r.h.s.\ using
a gluon condensate $\langle (\alpha_s/\pi) G^2\rangle = (0.36 \mbox{GeV})^2$
yields 3.32 GeV$^4$, so there is consistency to better than 2\%. The second
moment ($N=2$) involves uncertain four-quark condensates and is more
sensitive to the detailed form of the spectrum at higher energies. Its
discussion will be delegated to a forthcoming paper, but at this point
we can already conclude that the usual factorization ansatz for the four-quark
condensate assuming ground state dominance turns out not to be justified:
factorization underestimates the four-quark condensate by a large amount.
The statements about the low ($N=0,1$) spectral moments
are of course free of such uncertainties.

\begin{figure}[t]
\centerline{
\includegraphics[width=0.6\textwidth,angle=-90]{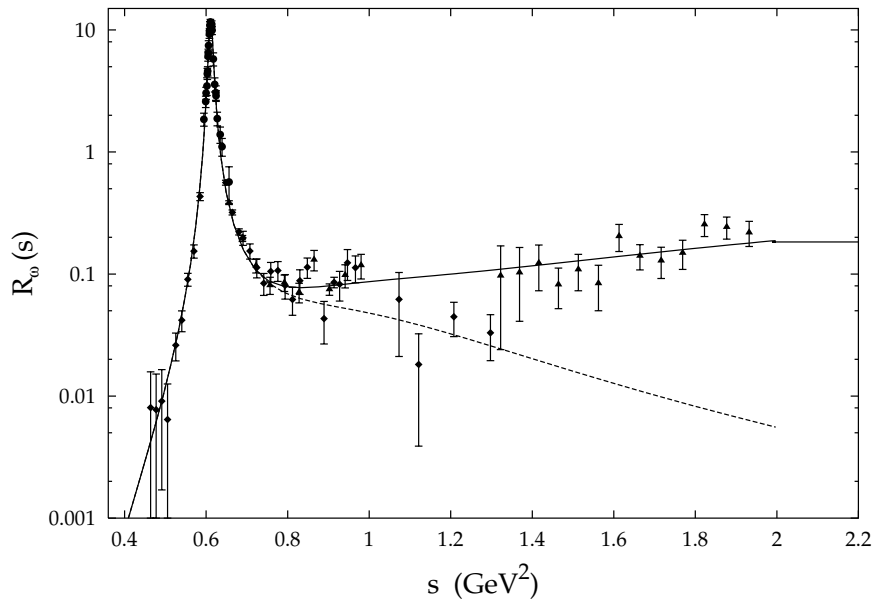}
}
\caption{Spectral distribution $R_{\omega}(s)$ in the isoscalar
($\omega$ meson) channel. Dashed curve: resonant part; solid curve: sum
of resonance and continuum contributions.
Experimental data: dots \cite{Barkov};
diamonds and triangles \cite{Dolinsky}, from $e^+ e^- \rightarrow n \pi$
with $n$ odd.}
\label{fig2}
\end{figure}

For the $\omega$ meson spectrum we are again guided by the effective
Lagrangian approach (Ref.\ \cite{KlinglKaiser}). The resonant part is
parametrized as in Ref.\ \cite{Achasov}.
(We stay in the $u,d$-quark sector and therefore omit $\omega \phi$ mixing).
Otherwise we follow a scheme analogous to that for the $\rho$ meson. The
QCD continuum starts at $s_c = 2$ GeV$^2$ and the same (linear) interpolation
around $s= \Delta^2$ is added to the resonance tail (dashed curve in Fig.\ 2)
between 0.8 and 2 GeV$^2$. The result (Fig.\ 2) compares quite well with the
data in the $I=0$ $e^+ e^- \rightarrow$ hadrons channel, though with admittedly
poor statistics in the region above the $\omega$ resonance. For the lowest
spectral moment we find $\int_0^{s_c} ds \, R_{\omega}(s) = 0.3917$ GeV$^2$
as compared to 0.3912 GeV$^2$ from the r.h.s.\ of the $N=0$ sum rule 
(\ref{eq:moments}). The $N=1$ moment gives 
$\int_0^{s_c} ds \, s R_{\omega}(s) = 0.371$ GeV$^4$, in perfect agreement
with the r.h.s.\ value 0.369 GeV$^4$. Consistency of the second moment
requires the same four-quark condensate as observed for the $\rho$ meson,
substantially larger than the value suggested by factorization into
$\langle \bar{q}q\rangle^2$.

In summary, the degree of consistency found in this approach
is quite impressive, at least for the first two moments of the spectral
distributions. In particular, the crossover between $\rho$ and $\omega$
resonance regions and the asymptotic continuum, although smooth, is still
controlled by the chiral symmetry breaking scale $\Delta = 4 \pi f_{\pi}$.

With this observation in mind we can briefly comment on the in-medium version
of the schematic model (\ref{eq:Rv}), with $m_V$ replaced by density dependent
masses $m_V^* (\rho)$ (for $V= \rho, \omega$) and the continuum onset
$s_0$ replaced by $s_0^*(\rho)$. Using such a parametrization Hatsuda and
Lee found $m_V^* \simeq m_V (1 - 0.16 \rho/\rho_0)$ in their in-medium
QCD sum rule analysis \cite{Hatsuda}. This result can now be given a
straightforward interpretation. The nuclear matter analogue of the sum rules
(\ref{eq:VMD0th}, \ref{eq:VMD1st}) gives the leading behaviour
$m_V^* (\rho) \simeq \sqrt{8} \pi f_{\pi}^* (\rho)$, where $f_{\pi}^*$ is
the pion decay constant in matter related to the time-component
of the axial current. The in-medium PCAC relation
$f_{\pi}^{*2} = (m_q / m_{\pi}^2) \langle \bar{q} q \rangle^* + \cdots$
implies that $f_{\pi}^*$ scales like the square root of the
quark condensate to leading chiral order. The magnitude of this
condensate at nuclear matter density $\rho_0$ (see \cite{Weise} for
a review and further references) is expected to drop
by about $1/3$ of its value at $\rho = 0$, so $m_V^*(\rho_0)/m_V \simeq
f_{\pi}^*(\rho_0)/f_{\pi}\simeq 5/6$ in the zero width limit. The same
result is also found in ref.\ \cite{KlinglKaiser} and in Brown-Rho scaling
\cite{BrownRho}. The $\omega$ meson which is predicted still to survive as a 
reasonably narrow state in nuclear matter \cite{KlinglKaiser}, should be
a good indicator of the way in which the chiral gap decreases with
increasing density. The much broader $\rho$ meson, on the other hand,
is probably not very useful for this purpose \cite{KlinglKaiser}.

In conclusion, introducing the chiral gap $\Delta = 4 \pi f_{\pi}$ as
a relevant scale in QCD spectral sum rules for vector mesons establishes
important connections with current algebra and the in-medium scaling of vector
meson masses. Further work along these lines is in progress.


\begin{thebibliography}{99}
\bibitem{SVZ} M.A. Shifman, A.I Vainshtein and Z.I. Zakharov, Nucl.\
Phys.\ B 147 (1979) 385, 448.
\bibitem{Chetyrkin}
 K.G. Chetyrkin, N.V. Krasnikov and A.N. Tavkhelidze, Phys.\ Lett.\ B 76 (1978)
 83.
\bibitem{Fischer}
 J. Fischer and P. Kol\'ar, Z. Phys. C 34 (1987) 375.
\bibitem{Klingl}
 F. Klingl and W. Weise, Eur. Phys. J. A 4 (1999) 225.
\bibitem{Hatsuda} T. Hatsuda and S.H. Lee, Phys.\ Rev.\ C 46 (1992) 34.
\bibitem{KlinglKaiser}
 F. Klingl, N. Kaiser and W. Weise, Nucl. Phys. A 424 (1997) 527.
\bibitem{Gimenez}
 V. Gim\'enez, J. Bordes and J. Pe\~narrocha, Nucl. Phys. B 357 (1991) 3.
\bibitem{Gorishny}
 S.G. Gorishny, A.L. Kataev and S.A. Larin, JETP Lett. 53 (1991) 127.
\bibitem{Pich} A. Pich, Nucl. Phys. B (Proc.\ Suppl.) 39 B (1995) 326.
\bibitem{KSFR} K. Kawarabayashi and M. Suzuki, Phys. Rev. Lett. 16 (1996) 255;
 Riazuddin and Fayyazuddin, Phys. Rev. 147 (1966) 1071.
\bibitem{Benayoun} M. Benayoun et al., Z. Phys. C 58 (1993) 31.
\bibitem{Barkov} L.M. Barkov et al., Nucl. Phys. B 256 (1985) 365.
\bibitem{Dolinsky} S.I. Dolinsky et al., Phys. Rep. 202 (1991) 99.
\bibitem{Bacci} C. Bacci et al., Nucl. Phys. B 184 (1981) 31.
\bibitem{Achasov} N.N. Achasov and A.A. Kozhevnikov, Phys. Rev. C 57
 (1998) 4334.
\bibitem{Weise} W. Weise, Les Houches Lectures, Session LXVI in:
 Trends in Nuclear Physics, 100 Years Later, Elsevier, Amsterdam (1998), p.\
 423.
\bibitem{BrownRho} G.E. Brown and M. Rho, Phys.\ Rev.\ Lett.\ 66 (1991) 2720.
\end{thebibliography}
\end{document}